\documentclass[final,5p,times,twocolumn,compress]{elsarticle}

\usepackage{graphicx}

\usepackage{amssymb}
\usepackage{amsthm}
\usepackage[fleqn]{amsmath}
\usepackage{bm}
\usepackage{array}
\usepackage{hyperref}
\usepackage{slashed}
\usepackage{braket}
\usepackage{booktabs}
\usepackage{multirow}
\usepackage{setspace}

\usepackage{cases}

\usepackage{xcolor}
\usepackage{tipa}
\usepackage{ulem}

\hypersetup{
    colorlinks=true,
    linkcolor=blue,
    filecolor=blue,
    urlcolor=blue,
    citecolor=blue,
}

\journal{Fundamental Research (2022) 
\href{https://doi.org/10.1016/j.fmre.2022.05.022}{doi:10.1016/j.fmre.2022.05.022}}

\begin{document}

\begin{frontmatter}



\title{Ultra-high energy cosmic neutrinos from gamma-ray bursts} 


\author[PKU]{Yanqi Huang}

\author[PKU,CHEP,CICQM]{Bo-Qiang Ma}
\ead{mabq@pku.edu.cn}

\address[PKU]{School of Physics, Peking University, Beijing 100871, China}
\address[CHEP]{Center for High Energy Physics, Peking University, Beijing 100871, China}
\address[CICQM]{Collaborative Innovation Center of Quantum Matter, Beijing, China}

\begin{abstract}
Based on recent proposal to associate IceCube TeV and PeV neutrino events with gamma-ray bursts~(GRBs)
by considering the Lorentz violation of neutrinos,
we provide a new estimate on the GRB neutrino flux and such result is much bigger than previous results by the IceCube Collaboration. Among these 24 neutrino ``shower" events above 60~TeV, 12 events are associated with GRBs. Such result is comparable with the prediction from GRB fireball models.
Analysis of track events provide consistent result with the shower events to associate high energy cosmic neutrinos with GRBs under the same Lorentz violation features of neutrinos. We also make a background estimation and reveal GRBs as a significant source for the ultra-high energy IceCube neutrino events.
Our work supports the Lorentz violation and $CPT$-violation of neutrinos, indicating new physics beyond relativity.
\end{abstract}


\end{frontmatter}


\section{Introduction}
Cosmogenic ultra-high energy neutrinos are important to reveal new features of the Universe. After around ten 
years of
measurements, the IceCube Collaboration has observed plenty of neutrino events with energies above 30~TeV, including a few PeV scale events~\cite{Aartsen:2013bka,Aartsen:2013jdh,Aartsen:2014gkd,Kopper:2015vzf,Aartsen:2016ngq,Aartsen:2018vtx}. However, the source of these ultra-high energy cosmic neutrinos still remains obscure. The neutrino emission associated with gamma-ray burst~(GRB) was proposed from the fireball model of GRBs~\cite{Eichler:1989ve,Paczynski:1994uv,Waxman:1997ti,Piran:1999bk}. By associating GRBs and some lower energy neutrinos with time difference of several hundred seconds, the IceCube Collaboration~\cite{Aartsen:2014aqy,Aartsen:2016qcr,Aartsen:2017wea} proposed constraints on GRB neutrinos with a small flux to exclude some fireball models~\cite{Waxman:1997ti,Ahlers:2011jj}.

However, because of the ultra-high energy and the long propagating distance between the GRB source and the detector, a neutrino detected months around the GRB trigger time might be 
associated with the GRB {by taking into account the Lorentz violation~(LV) effect, since the small residual effects of LV can be accumulated into observable effects among the propagation of several billion light-years}~\cite{Jacob:2006gn}. Based on the IceCube data, Amelino-Camelia and collaborators associated TeV scale neutrino events with GRB candidates within longer time range of three days~\cite{Amelino-Camelia:2015nqa,Amelino-Camelia:2016fuh,Amelino-Camelia:2016ohi}. It is also found in a recent study~\cite{Huang:2018} that all four events of PeV scale neutrinos observed by the IceCube observatory are associated with GRBs, by extending the temporal window with a longer time range of three months.
	The above associations of neutrino events with GRBs are obtained from the assumption of Lorentz violation of neutrinos and anti-neutrinos~\cite{Amelino-Camelia:2016ohi,Huang:2018} with also
	an asymmetry between neutrinos and anti-neutrinos to explain the existence of both time ``delay'' and ``advance'' events~\cite{Huang:2018,Zhang:2018otj}.
Therefore it is necessary to re-evaluate the GRB neutrino flux from the new associations of IceCube TeV and PeV neutrino events with GRBs~\cite{Amelino-Camelia:2016ohi,Huang:2018}.
In this work, we provide a new constraint on the GRB neutrino flux based on the suggested GRBs associated with the IceCube TeV and PeV events~\cite{Amelino-Camelia:2016ohi,Huang:2018} observed during 2010 to 2014~\cite{Website}. Among these 24 neutrino ``shower'' events above 60~TeV, 12 events are associated with GRBs~\cite{Amelino-Camelia:2016ohi,Huang:2018}. We calculate the differential limit of GRB neutrino flux and find it comparable with the prediction from GRB fireball models.
We also perform the analysis on the ``track'' events of IceCube neutrinos with energies higher than 30~TeV and find consistent result with the ``shower'' events under
the same Lorentz violation features of neutrinos.
We therefore reveal GRBs as a significant source for ultra-high energy cosmic neutrinos
	provided that there are Lorentz violation of neutrinos and anti-neutrinos~\cite{Amelino-Camelia:2016ohi,Huang:2018} with opposite signs to imply also an asymmetry between neutrinos and anti-neutrinos~\cite{Huang:2018,Zhang:2018otj}.

\section{GRB Neutrino Flux}
The phenomenological observations of GRBs can be well described by a relativistically expanding fireball of electrons, photons, and protons~\cite{Paczynski:1986px,Goodman:1986az,Waxman:1995vg}. Ultra-high energy neutrinos can be emitted during the fireball expansion~\cite{Eichler:1989ve,Paczynski:1994uv,Waxman:1997ti,Piran:1999bk}. From the viewpoint of the fireball model, the accelerated protons can scatter with the intense $\gamma$-ray background within the GRB fireball, and generate pions:
\begin{equation}
p+\gamma\to \Delta^+\to n+\pi^+.
\end{equation}
The charged pions and their muon daughters generate ultra-high energy cosmic neutrinos by the decay chains
\begin{equation}
\pi^+ \to \mu^+ +\nu_\mu \to e^+ +\nu_\mu+\bar\nu_\mu+\nu_e.
\end{equation}
The flux of ultra-high energy neutrinos is coincident with $\gamma$-rays, and the energy of produced neutrinos can be up to a few PeVs.
After long distance propagation in cosmic space, these
high energy $\nu_\mu$ and $\bar\nu_\mu$ neutrinos can interact with water or ice through charged-current interactions to produce high energy muons that manifest as extended Cherenkov light patterns in ice and can be detected by the IceCube detector as ``track'' events. Since the cosmic-ray-induced muon background is hard to remove, the Southern Hemisphere bursts are often excluded. To improve the sensitivity, another low-background channel named ``shower''	is introduced~\cite{IceCube document}. ``Shower'' events include cascades from $\nu_e$, $\nu_\tau$ charged-current interactions and all-flavor neutral-current interactions. Charged-current cascades include contributions from the electron~(or tau decay products), as well as the hadronic shower from the scattered parton.

The event rate detected on the earth can be described by the integration~\cite{Aartsen:2017wea}
\begin{equation}
\dot{N}=\int_\Omega d\Omega' \int dE_\nu A_{\rm eff}(E_\nu, \Omega')\times \Phi_\nu(E_\nu, \Omega'),
\label{integration}
\end{equation}
where $\dot{N}$ is the rate of neutrino events, $\Omega$ is the solid angle, $E_\nu$ is the neutrino energy, $A_{\rm eff}(E_\nu, \Omega')$ is the effective area of the IceCube detector and  $\Phi_\nu(E_\nu, \Omega')$ is the signal neutrino flux. The effective area $A_{\rm eff}(E_\nu, \Omega')$ for full-sky shower-like event searches with the 79-string IceCube detector is provided by the IceCube Collaboration~\cite{Aartsen:2017wea}. In the case of null observation, the quasi-differential limit of the neutrino flux can be written as~\cite{Anchordoqui:2002vb,Aartsen:2018vtx}
\begin{equation}
\Phi_\nu(E_\nu)=3\frac{N}{4\pi E_\nu T \log 10 \sum A_{\rm eff}(E_\nu)},
\label{differential limit}
\end{equation}
where $N$ is the event number, and $T$ is the observation time. The summation includes effective areas of all three neutrino flavors.
An equal flavor ratio of
neutrino fluxes $\nu_e : \nu_{\mu}: \nu_{\tau}= 1 : 1 : 1$ at the Earth is
assumed under the standard neutrino oscillation scheme, though both neutrino oscillations and neutrino decays may make the neutrino propagation in the Universe more complicated~\cite{Huang-Ma}.
In our discussion, we use the four years IceCube data, and the observation time $T=1347$ days~\cite{Website}.

\section{IceCube Shower Events}

\begin{table*}[]
	\caption{Properties of 12 shower events. The 12 GRB candidates are suggested from the associated GRBs of IceCube neutrinos with energies above 60~TeV by the maximum correlation criterion~\cite{Amelino-Camelia:2016fuh,Amelino-Camelia:2016ohi,Huang:2018}. The event serial numbers here are provided by the IceCube database.
		Only ``shower" events from refs.~\cite{Amelino-Camelia:2016fuh,Amelino-Camelia:2016ohi,Huang:2018} are listed and a 2.6~PeV ``track" event in ref.~\cite{Huang:2018} will be discussed in next section.
		The mark $^*$ represents the estimated value of the redshift. $N_{\rm B}$ is the estimated background GRB number of each event, and $N'_{\rm B}$ is another option of the estimated background GRB number with fixed GRB rate. }
	\label{tab:1}
	\centering
	\begin{tabular*}{0.9\textwidth}{@{\extracolsep\fill}ccccccc}
		\noalign{\vspace{0.5ex}}		
		\hline
		\hline
		\noalign{\vspace{0.5ex}}
		event &	GRB & $z$ & $\Delta t_{\rm obs}~(10^3~\rm s)$ & $E$~(TeV) & $N_B$ &  $N'_B$	\\
		\noalign{\vspace{0.5ex}}
		\hline
		\noalign{\vspace{0.5ex}}
		$\#$2& 100605A&1.497$^*$&-113.051&117.0 &0.024 &  0.019 \\
		\noalign{\vspace{0.5ex}}
		$\#$9&  110503A&1.613&80.335&63.2& 0.083 &  0.07\\
		\noalign{\vspace{0.5ex}}
		$\#$11& 110531A&1.497$^*$&185.146&88.4& 0.36 &  0.47\\
		\noalign{\vspace{0.5ex}}
		$\#$12& 110625B&1.497$^*$&160.909&104.1&0.06 &  0.13\\
		\noalign{\vspace{0.5ex}}
		$\#$14& 110725A&2.15$^*$&1320.217&1040.7& 0.02 &  0.05 \\
		\noalign{\vspace{0.5ex}}
		$\#$19&  111229A&1.3805&73.960&71.5& 0.008 &  0.004\\
		\noalign{\vspace{0.5ex}}
		$\#$20& 120119C&2.15$^*$&-1940.176&1140.8& 1.8  &  4.1\\
		\noalign{\vspace{0.5ex}}
		$\#$26& 120219A&1.497$^*$&229.039&210.0& 0.22 &  0.22\\
		\noalign{\vspace{0.5ex}}
		$\#$33& 121023A&0.6$^*$&-171.072&384.7& 0.19 &  0.14\\
		\noalign{\vspace{0.5ex}}
		$\#$35& 130121A&2.15$^*$&-2091.621&2003.7& 0.05 &  0.14\\
		\noalign{\vspace{0.5ex}}		
		$\#$40&	130730A&1.497$^*$&-179.641&157.3& 0.069  &  0.05\\	
		\noalign{\vspace{0.5ex}}
		$\#$42& 131118A&1.497$^*$&-146.960&76.3& 0.88  &  0.91\\
		\noalign{\vspace{0.5ex}}
		\hline
		\hline
		
	\end{tabular*}
	
\end{table*}

From 2010 to 2014, the IceCube detector observed 32 neutrino events with energies above 60 TeV, and 24 of them are ``shower'' events. To select probable GRB neutrinos, we use time and direction criteria to find the associated GRB for each neutrino event~\cite{Amelino-Camelia:2016fuh,Amelino-Camelia:2016ohi,Huang:2018}. The time mismatch could be extended by the ultra-high energy and the long propagation distance of neutrinos. So it is reasonable to expand the time range with the increase of energy. For the TeV neutrinos, a time window of three days is adopted~\cite{Amelino-Camelia:2016ohi}. For the two 1~PeV events, we include the GRBs detected within one month before or after the neutrinos. For the 2~PeV event, the time window is two months~\cite{Huang:2018}. By adopting these time windows, we find that each IceCube neutrino event may have more than one GRB candidates. To choose the most likely associated GRBs, we use the maximum correlation criterion~\cite{Amelino-Camelia:2016fuh,Amelino-Camelia:2016ohi,Huang:2018} and the strict time criterion that requires~\cite{Huang:2018}
\begin{equation}
|\frac{\Delta t_{\rm obs}}{1+z}-s\cdot\frac{K}{E_{\rm LV}}|<30\%\cdot\frac{K}{E_{\rm LV}}\label{strict time},
\end{equation}
where $E_{\rm LV}$ is the Lorentz violation scale, $K$ is the LV factor
	and $s=\pm1$ is the sign factor of the LV correction term. Since the linear LV correction implies the $CPT$-odd term in an effective field theory, neutrinos and anti-neutrinos have opposite sign factors, i.e.,
	there is an asymmetry between neutrinos and anti-neutrinos~\cite{Huang:2018,Zhang:2018otj}.
This time criterion requires that the observed time differences between the IceCube event and its associated GRB satisfy the possible Lorentz violation regularity
proposed from GRB photons, TeV and PeV neutrinos, with the consideration of errors.
The maximum correlation criterion and the strict time criterion can effectively
depress the effects of GRB backgrounds. While considering the backgrounds, one should
focus on the refined strict time criterion, rather than a rough time window of
months.
For the directional criterion, a two dimensional circular Gaussian~\cite{Amelino-Camelia:2016fuh}
\begin{equation}
P(\nu, \mathrm{GRB})=\frac{1}{2\pi\sigma^2}\exp(-\frac{\Delta \Psi^2}{2\sigma^2}),
\end{equation}
is introduced, where $\Delta \Psi$ is the angular separation between GRB and neutrino, and $\sigma=\sqrt{\sigma_{\rm GRB}^2+\sigma_{\nu}^2}$ is the standard deviation based on the angular uncertainties of GRB and neutrino measurements. {
	The detailed values of both $\Delta \Psi$ and $\sigma$ can be found in the previous studies~\cite{Amelino-Camelia:2016fuh,Amelino-Camelia:2016ohi,Huang:2018}.} In our analysis, we consider the GRBs whose angular separation is smaller than $3\sigma$.
For the redshift, some of the GRB candidates do not have determined redshift yet. As discussed in previous studies~\cite{Amelino-Camelia:2016fuh,Amelino-Camelia:2016ohi,Huang:2018}, here we use a most likely estimated value obtained as the average value of all known redshifts that have been measured so far. What needs to be emphasized is that a relatively wide error range of $0.5z\sim2z$ is considered for the unknown redshifts in our analysis. Such a range covers most of the known redshifts of GRBs detected so far and can also separate the ``long burst" and the ``short burst", since the two error ranges have no overlap. In our analysis, the error of redshift plays a role in the errors of fitting parameters.
Among the 24 ``shower'' events above 60~TeV, 12 events can be associated with GRBs. The properties of probable GRB neutrinos that satisfy both time and direction criteria are listed in Table~\ref{tab:1}.

We estimate the statistical significance of these 12 associated GRBs by calculating the background GRB number $N_{\rm B}$, which is defined as
\begin{equation}
N_{\rm B}=\frac{1}{4\pi}N_{\rm T} \times \Delta\Omega,
\end{equation}
where $\Delta \Omega$ is the space angle obtained from the angular separation $\Delta \Psi$, and $N_{\rm T}$ is the total number of GRBs that satisfy the time criterion Eq.~(\ref{strict time}) of neutrino events. Here we count all GRBs collected in the GRB catalog~\cite{GRB database}.
	The localization information of the GRBs that associated with 9 shower events with the energy from 60 to 500 TeV can be found in
	refs.~\cite{Amelino-Camelia:2016fuh,Amelino-Camelia:2016ohi}. The directional information of 4 events with the
	energy above PeV can be found in ref.~\cite{Huang:2018}. Actually the detail
	information of all these GRBs can be found in the GRB catalogs
	that are collected on the website~\cite{GRB database} http://www.mpe.mpg.de/~jcg/grblink.html, so as the GRBs associated with track events which are proposed in this work.
The background estimations are also listed in Table~\ref{tab:1}. We find that most of GRB candidates stand out from the background, and only 2 candidates have background errors larger than 50\%. The sum of $N_{\rm B}$ is 3.76, which is less than 30\% of the total  number. Furthermore, if regarding the 2 large background candidates as the statistical errors and excluding them, we can get the total background GRB number $\Sigma'N_{\rm B}=1.084$ for the rest 10 candidates. Therefore, these 10 candidates are approximately 10 times higher than the background value.

To reconfirm the background estimation of associated GRBs, we adopt another option to count the total number of GRBs $N_{\rm T}$ with a fixed GRB rate. Here we assume that there are 667 GRBs over the full sky per year, which is the same as the GRB rate used in the previous publications of IceCube collaboration~\cite{Abbasi:2011qc}. The results are also listed in Table~\ref{tab:1} as $N'_{\rm B}$. We can also find that most of the GRB candidates stand out from the background, and $N'_{\rm B}$ obtained from the fixed GRB rate are almost the same as $N_{\rm B}$ obtained from the GRB database. The sum of $N'_{\rm B}$ is 6.303. If regarding the 2 large background candidates as errors and excluding them, the total background GRB number is $\Sigma'N'_{\rm B}$=1.293 of the rest 10 candidates with the fixed GRB rate. Therefore, the number of GRB candidates that can be associated with IceCube shower events are much higher than the GRB backgrounds, whether we count the GRBs in the time window or we estimate $N'_{\rm B5}$ with a fixed GRB rate.

\begin{figure}[]
	

	\centering
	\includegraphics[width=\linewidth]{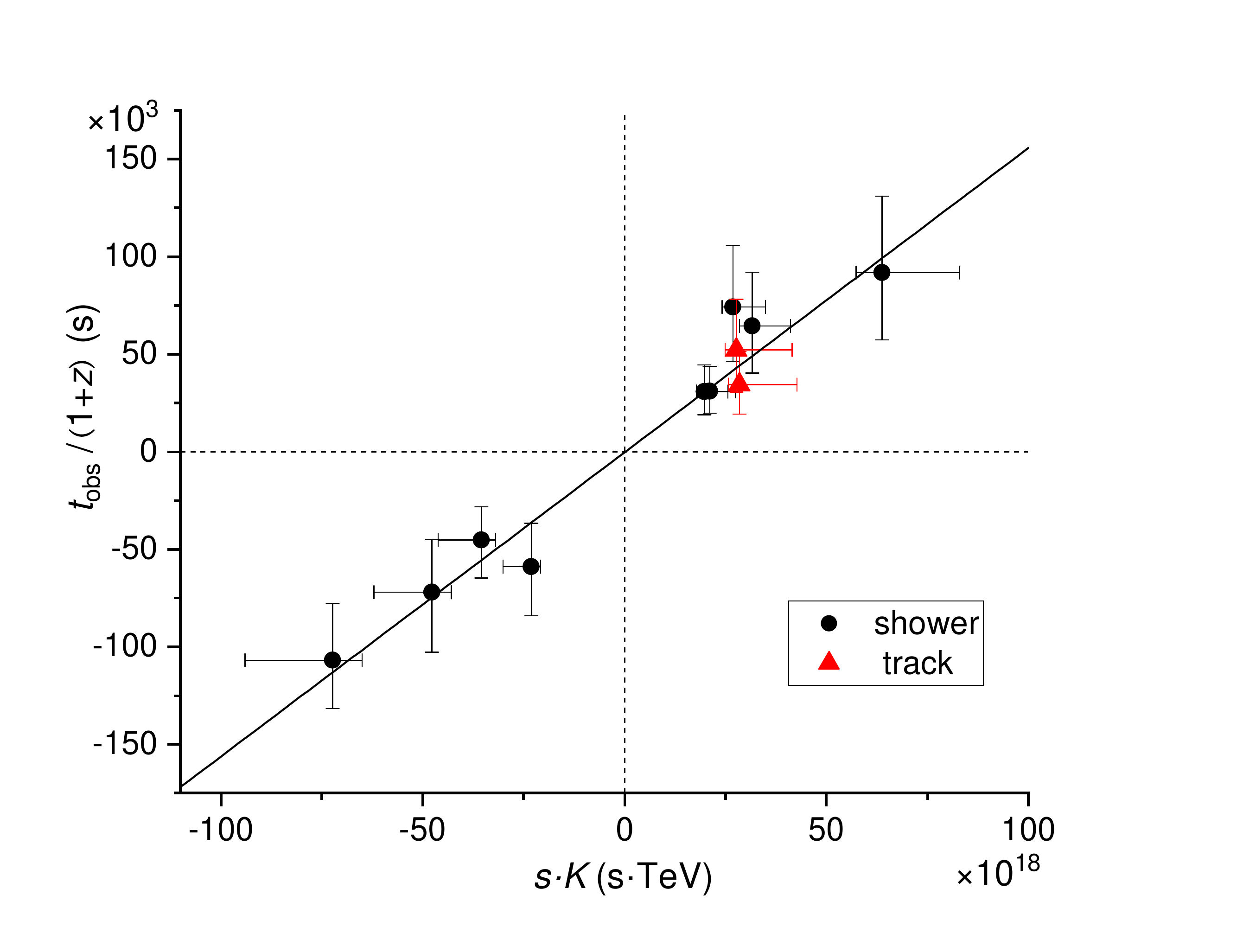}

	\caption{{Accordance between TeV scale track and shower events that can be associated with GRB candidates.} Nine black points are TeV shower events, two red triangles are TeV track events $\#23$ and $\#44$. The solid line represents the regularity of energy dependent speed variation found in both TeV and PeV shower events~\cite{Huang:2018}. The error bars here are estimated according to the same method in the reference~\cite{Huang:2018}. We can find that the two track events are in good accordance with the regularity.}
	\label{fig2}
\end{figure}

	One feature to be noticed is that from Fig.~\ref{fig2} and Fig.~\ref{fig3}, there exist both time ``delay'' and ``advance'' events, i.e., some neutrino events are found to be
	later than the GRB photons whereas some neutrino events are found to be earlier than the photons. One possibility is that the associations of neutrino
	events with GRBs are purely accidental. However, it is suggested in ref.~\cite{Huang:2018} that the IceCube Neutrino Observatory can not distinguish between neutrinos and anti-neutrinos, therefore the co-existence of both time ``delay'' and ``advance'' events can be explained by the different propagation properties between neutrinos and anti-neutrinos. The different propagation properties between neutrinos and anti-neutrinos can be explained by the $CPT$-odd feature of the linear Lorentz violation~\cite{Zhang:2018otj}, indicating an asymmetry between neutrinos and anti-neutrinos.

\section{IceCube Track Events}

\begin{table*}[]
	\caption{Properties of 12 track events. The 12 probable GRB neutrinos are based on track events of the 2010 to 2014 IceCube data~\cite{Website}. $E$ is the deposited energy of the track event, which is the lower bound of the neutrino energy. In the top part of the table, the events marked by $^a$ has a 50\% upper error of energy ~\cite{Amelino-Camelia:2016fuh,Amelino-Camelia:2016ohi,Huang:2018}, and the event $\#13$ marked by $^b$ has an upper error which is 3 times of the energy. In the bottom part of the table, the estimated total energy of neutrinos is up to 2 PeV~\cite{Huang:2018}. The event serial numbers here are provided by the IceCube database, {
			except the $\#7856$ which is the ATel ID of the GRB 140427A}.
		The mark $^*$ represents the estimated value of the redshift. $N_{\rm B}$ is the estimated background GRB number of each event, and $N'_{\rm B}$ is another option of the estimated background GRB number with fixed GRB rate. }
	\label{tab:2}
	\centering
	\begin{tabular*}{0.9\textwidth}{@{\extracolsep\fill}ccccccccc}
		\noalign{\vspace{0.5ex}}		
		\hline
		\hline
		\noalign{\vspace{0.5ex}}
		event &	GRB & $z$ & $\Delta t_{\rm obs}~(10^3\rm s)$ & $E$~(TeV) &{$\Delta \Psi~(^\circ)$}&{$\sigma~(^\circ)$}& $N_B$& $N'_B$\\
		\noalign{\vspace{0.5ex}}
		\hline
		\noalign{\vspace{0.5ex}}
		$\#$13$^b$& 110629A$^b$&2.15$^*$&1290.76&252.7&5.7&5.3&0.04 &  0.07 \\
		\noalign{\vspace{0.5ex}}
		$\#$23$^a$&  120121C$^a$&2.15$^*$&164.343&82.2&10.4&4.2& 0.04 &  0.03 \\
		\noalign{\vspace{0.5ex}}
		$\#$44$^a$& 140113B$^a$&2.15$^*$&108.346&84.6&16.9&12.6& 0.13 &  0.05 \\
		\noalign{\vspace{0.5ex}}
		$\#7856^a$& 140427A$^a$&2.15$^*$&3827.439&$2.6\times10^3$&25.8&23.4&1.14 &   4.03  \\
		\noalign{\vspace{0.5ex}}
		\hline
		\noalign{\vspace{0.5ex}}
		$\#$3& 100805C&2.15$^*$&3216.32& $78.7$ &15.3&5.2&0.38 &  1.21  \\
		\noalign{\vspace{0.5ex}}
		$\#$5&  101112B&2.15$^*$&-37.3568&71.4&10.8&6.0& 0.04 &  0.007  \\
		\noalign{\vspace{0.5ex}}
		$\#$8& 101214B&2.15$^*$&5428.26&32.6&9.9&6.6& 0.33 &  0.86  \\
		\noalign{\vspace{0.5ex}}
		$\#$18& 120102B&2.15$^*$&341.15&31.5&4.7&4.9& 0.01 &  0.015 \\
		\noalign{\vspace{0.5ex}}
		$\#$28& 120522A&2.15$^*$&-1776.58&46.1&9.2&3.3& 0.09 &  0.24  \\
		\noalign{\vspace{0.5ex}}
		$\#$37& 130427A& 0.34 &-1653.37&30.8&9.0&3.2&0.08 &  0.22\\
		\noalign{\vspace{0.5ex}}		
		$\#$38& 130521A&2.15$^*$&3037.78&200.5&5.7&3.2& 0.08 &  0.16\\
		\noalign{\vspace{0.5ex}}
		$\#$47&140105A&0.5$^*$&3675.07&74.3&15.1&6.9& 0.48 &  1.38  \\
		\noalign{\vspace{0.5ex}}
		\hline
		\hline
		
	\end{tabular*}
	
\end{table*}

\begin{figure}[]
	\centering
	\includegraphics[width=\linewidth]{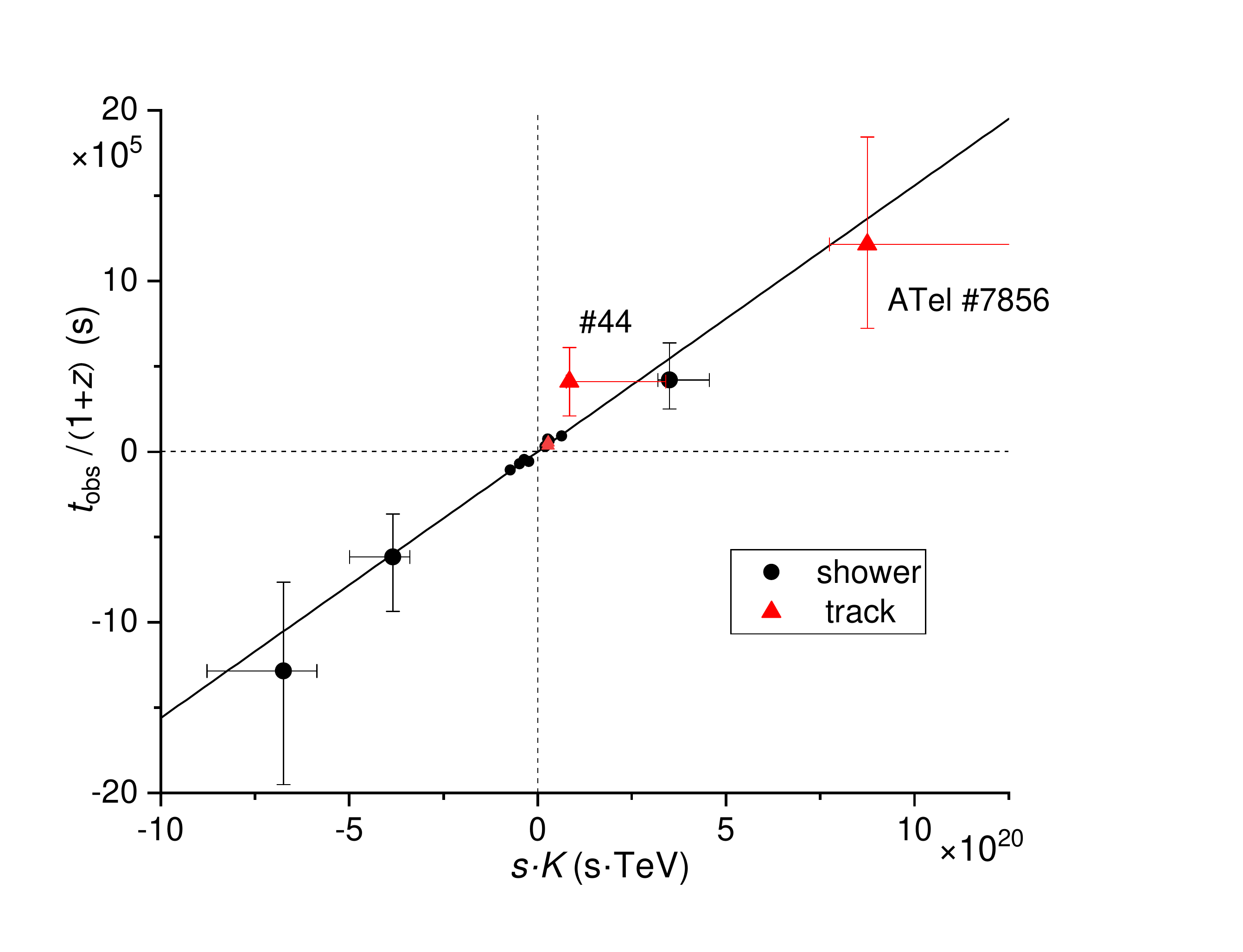}
	\caption{{Accordance between track and shower events that can be associated with GRB candidates.} Twelve black points are shower events, and three of them are PeV events with error bars. Four red triangles are track events, {
			with two events without error bars are overlapped together and}
		two ones with error bars are the event $\#13$ and ATel $7856$. The solid line represents the regularity of energy dependent speed variation found in both TeV and PeV shower events~\cite{Huang:2018}. The error bars of event ATel $7856$ are estimated according to the same method in the reference~\cite{Huang:2018}, but the event $\#13$ has an upper energy error which is 3 times of the energy. We can also find accordance between shower and track events if considering error bars.We can find that the two track events are in good accordance with the regularity.}
	\label{fig3}
	
\end{figure}

The IceCube ``track'' events are regarded as high energy muons produced in the charged-current interaction between high energy $\nu_\mu$~(or $\bar{\nu}_\mu$) and ice.
From 2010 to 2014~\cite{Website}, the IceCube detector observed 14 track events with energies above 30~TeV, and 8 of them are above 60~TeV. Together with event ATel $\#7856$ with the energy of 2.6~PeV observed a little latter in June 2014, 15 track events are taken into consideration in our analysis.

Though the muon track seems to have high directional resolution, the angular difference between muon and neutrino is not negligible.  The track event is regarded as the high
energy muon produced by the charged-current interaction of neutrino
with matter (i.e., actually the neutrino scattering on a parton
with the production of a charged lepton and a scattered parton
absorbing the charged $W^{\pm}$ boson). There are two parts of
final particles, the charged lepton and the cascade due to the
hadronization of the scattered parton. As the struck parton energy
is small, we can consider the process as the splitting of the
neutrino into a charged lepton and a scattered parton. So the
direction of the neutrino should be along the momentum sum of the
charged lepton and the cascade. The angular difference between leptons and cascades
is not negligible, so as that between leptons and neutrinos. A
recent data analysis from the IceCube Collaboration~\cite{Aartsen:2018vez} showed a 8.1 degrees mismatch between
cascade and track zenith angles, which indicates a similar scale of
angular difference between the muon track and the neutrino. As a conservative estimation,
we assume a 2 degrees angular difference between muon track and neutrino. Namely, in our analysis,
the angular uncertainty of neutrinos is 2 degrees larger than that of muon tracks, which are observed and reported as ``track'' events.

On the other hand, the ``deposited energy'' (the particle energy collected by
the IceCube detector) of the track event may be only a part of the
muon energy, since the produced vertex of muon tracks may be
located outside the instrumental volume~\cite{Aartsen:2014aqy}. The
muon energy is only a part of the total energy of the neutrino too.
Therefore, the deposited energy of track events may be much lower
than the true energy of neutrinos. So the neutrino energy can range
from the track energy to a large value of a few PeV in principle.
Such large energy uncertainties lead to up to months errors in
time windows.

To estimate the energy uncertainties, we consider three different levels. The first method is the same as our previous estimation~\cite{Huang:2018}, assuming that the positive error of energy is 50\% and the negative error is provided by IceCube measurements. In this method, we find that two TeV scale track events, i.e., event $\#23$ and $\#44$, and the 2.6~PeV event ATel $\#7856$ can be associated with GRBs and satisfy the regularity of energy dependent speed variation found in shower events~\cite{Huang:2018}. The accordance between these three track events and shower events in Table~\ref{tab:1} are shown in Fig.~\ref{fig2} and Fig.~\ref{fig3}. The properties of the three probable GRB neutrinos from track events are listed in Table~\ref{tab:2} with the mark $^a$. The detailed values of both
$\Delta
\Psi$ and $\sigma$ for the track events can be found in Table~\ref{tab:2}.

Since the energy of track events is just a part of the muon energy, and the muon energy is also just a part of the total energy of neutrinos, the real neutrino energy may be much higher than the energy of the track events. So as the second estimation method, we assume that the positive error is three times of the deposited energy, i.e., the energy detected in the instrumental volume is assumed to be only 25\% of the total neutrino energy. According to this error estimation, we find another track event $\#13$ can be also associated with GRB. The event $\#13$ is also shown in Fig.~\ref{fig3} and in Table~\ref{tab:2} with the mark $^b$.

Further more, since the PeV scale neutrinos are observed in both shower-like and track-like signals, we may reasonably assume that the total neutrino energy of track events can be up to PeV, though most of the energy are not detected. Such large energies lead to long time windows. Here we adopt a two months time window as our third estimation method, which is the same as the 2~PeV shower event. We find that in this method, another eight events can be also associated with GRBs, whose properties are listed in the bottom part of Table~\ref{tab:2}. Now we can see that, most of the track events~(12 of 15 events) can be associated with GRBs, if the large uncertainties of track events are completely considered.

To estimate the significance of the associated GRBs, we also calculate the background GRB number $N_B$ of track events in two options. The calculation methods are similar to that of shower events. The only difference is that the uncertainty of energies, as well as that of the LV factor $K$, is too large to restrict the time criterion. Therefore, the GRB total number $N_T$ here consists of GRBs whose observed time $T$ satisfy
\begin{equation}
|T-\Delta t_{\rm obs}|<30\% \cdot \Delta t_{\rm obs},
\label{time windows2}
\end{equation}
where $\Delta t_{\rm obs}$ is the observed time of the associated GRB of a track event. $N_B$ of each event can be obtained immediately, as listed in Table~\ref{tab:2}. The sum of $N_B$ is 2.89. If regarding the large background event ATel $\#7856$ as the statistical error and excluding it, we obtain $\Sigma'N_B=1.75$ for the 11 candidates of track events. Therefore, the associations between GRB candidates and the track events are also significant in comparison with the background.

In the fixed GRB rate case, the time window is also adjusted according to Eq.~(\ref{time windows2}), and the estimated GRB rate is 667 GRBs per year. The results of $N'_{\rm B}$ are listed in Table~\ref{tab:2}. The sum of $N'_{\rm B}$ is 8.272, and the $\Sigma'N'_{\rm B}$ of the 11 candidates except ATel $\# 7856$ is 4.242. So the number of GRB candidates that can be associated with track events is still larger than the GRB backgrounds. Considering the unclear uncertainties in direction and the large uncertainties in energy, it is understandable that backgrounds of track events are larger than that of shower events.

From above analysis of the track events, we find that 12 of 15 track events can be associated with GRBs
if taking the large errors in both time and direction into consideration. What is more, even if we do not take these errors into
consideration but use the same error estimation with shower events, we can still find 3 track events that can be
associated with GRB candidates and satisfy the same regularity found in TeV and PeV shower events~\cite{Amelino-Camelia:2016ohi,Huang:2018}.
Therefore our results of track events are consistent with the shower events to associate high energy cosmic neutrinos with GRBs with a same Lorentz violation scale $E_{\mathrm{LV}}$,
which is the only free parameter that can be determined by a single event in principle. One can easily finger out that the
significance of our result is very high considering that 12 shower events and also 12 track events fall on
a same line with a fixed $E_{\mathrm{LV}}$.
{
	It is interesting to notice that there are both time ``delay'' and ``advance'' events,
	as can be seen from Table~\ref{tab:2}, thus the track events also support the proposal~\cite{Huang:2018} of Lorentz violation with different propagation properties between neutrinos and anti-neutrinos.}

\section{Results}

\begin{figure}[]
	\centering
	\includegraphics[width=\linewidth]{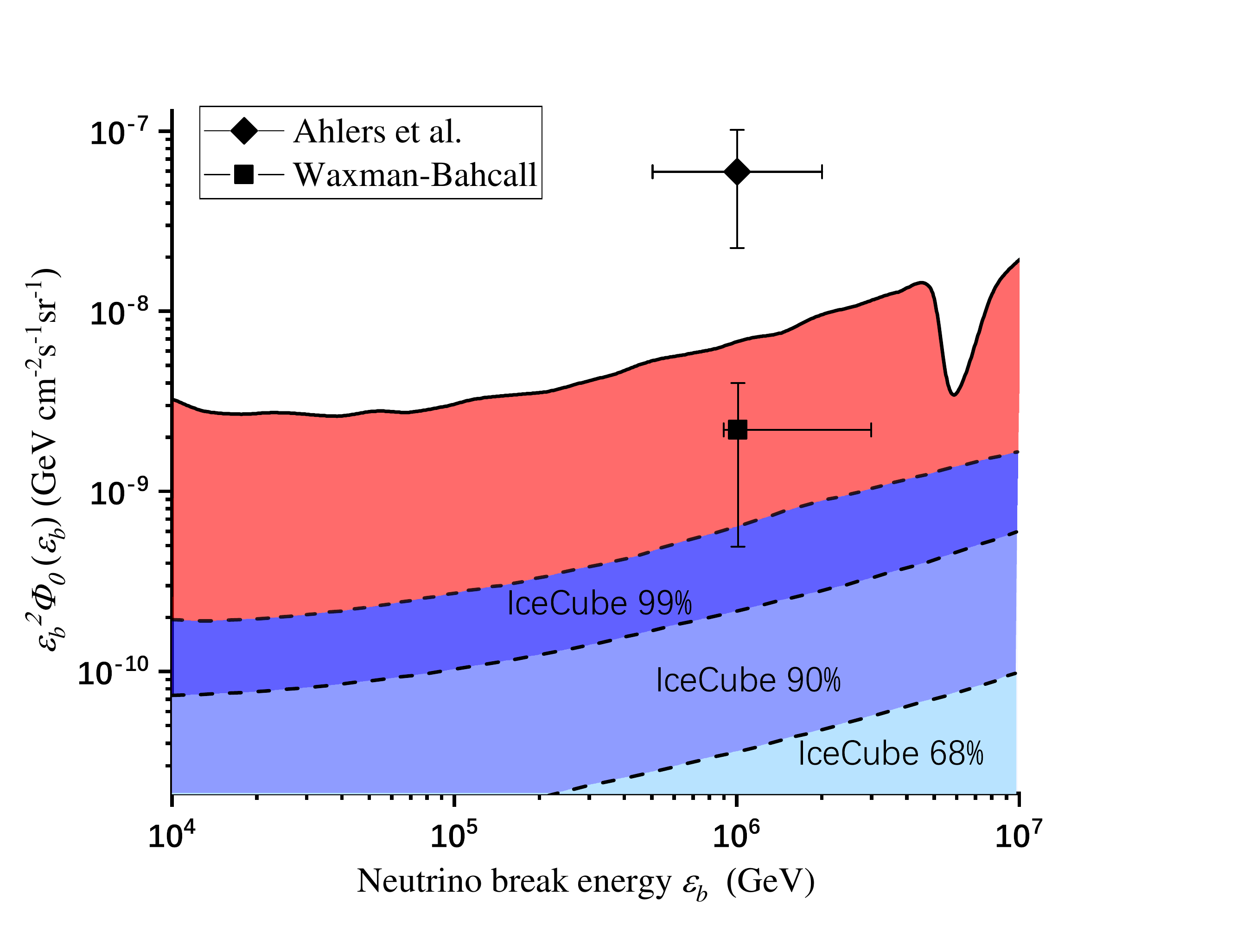}
	\caption{{Estimations on GRB neutrino flux based on full-sky shower-like IceCube neutrino events from 2010 to 2014.} The solid curve and the red region are our results. The dashed curves and the blue regions are the IceCube excluded regions for 99\%, 90\% and 68\% CL~\cite{Aartsen:2017wea}. The square and diamond points are predictions from the proton escape case~\cite{Waxman:1997ti} and the neutron escape case~\cite{Ahlers:2011jj} of the GRB fireball model. We can find that our new limit of GRB neutrino flux is compatible with the proton escape model, whereas two fireball models were excluded by the IceCube limit at 90\% CL~\cite{Aartsen:2017wea}.}
	\label{fig1}
	
\end{figure}

Here we calculate the theoretical GRB neutrino flux according to two models, the proton escape model~\cite{Waxman:1997ti} and the neutron escape model~\cite{Ahlers:2011jj}. We assume that there are 667 GRBs all over the full sky per year. The baryonic loading, which means the ratio of fireball energy in protons to electrons, is set as $f_p=10$. The bulk Lorentz factor $\Gamma$ of the fireball is set as a benchmark value $\Gamma=300$, which leads to a double broken power laws of neutrino spectra peaking around 100~TeV~\cite{Waxman:1997ti}. To compare the theoretical results with the measurements of the ultra-high energy neutrino flux, we introduce a generic double broken power-law neutrino flux~\cite{Aartsen:2014aqy}:
\begin{equation}
\Phi_\nu(E_\nu)=\Phi_0 \times
\left\lbrace
\begin{array}{ll}
\epsilon_b^{-1}E_\nu^{-1}, & E_\nu\le\epsilon_b;\\
E_\nu^{-2}, & \epsilon_b< E_\nu\le10\epsilon_b;\\
E_\nu^{-4}(10\epsilon_b)^2, & 10\epsilon_b< E_\nu,\\
\end{array}
\right.
\end{equation}
where $\epsilon_b$ is the first break energy and $\Phi_0$ is the quasi-diffuse spectral normalization flux. Considering Eq.~(\ref{differential limit}), we get the generic double broken power-law neutrino spectrum as shown in Fig.~\ref{fig1}. {As a most conservative estimation on order of magnitude, we only consider shower events in our analysis. Hence, it is acceptable that the GRB neutrino flux are two or three times larger than our estimation.} For comparison, the IceCube excluded regions for 68\%, 90\% and 99\% confidence level~(CL)~\cite{Aartsen:2017wea} are also shown in the plot. We can find from Fig.~\ref{fig1} that our new limit of GRB neutrino flux is compatible with the proton escape model, whereas two fireball models were claimed to have been excluded by the IceCube limit at 90\% CL~\cite{Aartsen:2017wea}.

In our discussion, IceCube ``shower'' events with energies above 60~TeV and ``track'' events with energies above 30~TeV are analyzed. 12 of 24 shower events and 12 of 15 track events can be associated with GRB candidates, which indicates that a part of these IceCube ultra-high energy neutrinos are emitted from the GRB source. The sources of ultra-high energy neutrinos might have more possibilities. As an example, some researches indicate that two ultra-high energy neutrinos might be emitted by blazars~\cite{Kadler:2016ygj,
IceCube:2018dnn,IceCube:2018cha}. Such a small number of neutrino events associated with blazars does not conflict with the results to attribute GRBs as a significant source for IceCube TeV and PeV events~\cite{Amelino-Camelia:2016ohi,Huang:2018}.

\section{Discussions}
\label{sec:Discussions}

We see that our result on the GRB neutrino flux differs significantly from that by the IceCube Collaboration~\cite{Aartsen:2016qcr,Aartsen:2017wea}. The reason for the difference is due to different choices of the time window between the
neutrino events and the associated GRBs.
In our analysis, we adopt a larger time range from a few days for TeV events~\cite{Amelino-Camelia:2016ohi} to a few months for PeV events~\cite{Huang:2018}
{
	by taking into account a
	sizable Lorentz violation effect of neutrinos and anti-neutrinos},
whereas the constraints on the GRB neutrino flux proposed by the IceCube Collaboration are based on an assumption that GRB neutrinos should be detected in very close temporal coincidence with the associated GRBs within a few hundred seconds~\cite{Aartsen:2016qcr,Aartsen:2017wea}.
In the combined linear fitting of TeV and PeV neutrinos~\cite{Huang:2018},
we also obtained an intrinsic time difference
{
	between neutrino and photon emissions}
as $\Delta t_{\rm in} = (1.7 \pm 3.6) \times 10^3 ~\rm s$.
Though we have little information about the intrinsic time difference,
$\Delta t_{\rm in}$ of the order of 1-2 hour can be safely neglected in the
fittings of TeV and PeV neutrinos, since it is much more shorter than the
observed time differences of the order of days or even up to months.

In fact, the GRB neutrinos are emitted from long distant sources, and speed variation may be caused by different reasons~\cite{Jacob:2006gn}. The {\textit{in vacuo}} dispersion due to Lorentz invariance violation is one of the probable options, as well as the cosmic matter effect. Therefore, the time difference between neutrino observed time and the GRB trigger time may be days or even months, depending on the neutrino energy {
	with the assumption of a sizable Lorentz violation of neutrinos}. On the other hand, ultra-high energy neutrinos can be distinguished from neutrinos produced in the atmosphere and other backgrounds.
No matter from the theoretical analyses~\cite{Jacob:2006gn,Amelino-Camelia:2015nqa,Amelino-Camelia:2016fuh,Amelino-Camelia:2016ohi} or the IceCube measurements~\cite{Aartsen:2013jdh,Aartsen:2014gkd,Aartsen:2016ngq,Aartsen:2018vtx,Kopper:2015vzf}, ultra-high energy neutrinos above 60~TeV can stand out from backgrounds, thus render our analysis on the limit of GRB neutrino flux more convincing.

The idea of using neutrinos from GRBs to explore quantum-gravity-induced
Lorentz violation was first proposed by Jacob and Piran in
2007~\cite{Jacob:2006gn}. Amelino-Camelia and collaborators found the
associations between GRBs and nine IceCube shower neutrino events
with the energies between 60 to 500 TeV~\cite{Amelino-Camelia:2016fuh,Amelino-Camelia:2016ohi}, and
found roughly compatible energy-dependent speed variation features
between GRB neutrinos and GRB photons. In our former work~\cite{Huang:2018}, we found that all four IceCube events
of PeV scale neutrinos
can be associated with GRBs.
	Furthermore, we also suggested different propagation properties
	between neutrinos and anti-neutrinos due to the existence of both time ``delay'' and ``advance'' events.
	Such different neutrino/anti-neutrino propagation properties can be described in an effective field theory framework with $CPT$-odd terms of Lorentz invariance violation~\cite{Zhang:2018otj}.
Therefore both PeV and TeV neutrino events
satisfy a same speed variation regularity
	based on the assumption of an observable Lorentz violation effect
	of neutrinos and anti-neutrinos.
Recently, from the associations between GRBs and near-TeV events reported by
IceCube Collaboration, it is found that 12 near-TeV northern
hemisphere track events can fall on the same line~\cite{Huang:2019}. In
this current paper, we show that another 3 or more track events above 30 TeV can
be associated with GRBs and satisfy the same regularity. Based on these
supports, the energy-dependent speed variation feature of
ultrahigh-energy neutrinos with the LV scale $E_{\rm
	LV}=(6.4\pm1.5)
\times10^{17}{\rm GeV}$
emergences as a more supportive regularity.


\section{Summary}
\label{sec:Summary}

In summary, we analyze the IceCube data in four years from 2010 to 2014 and get a new constraint on the GRB neutrino flux. Among all 24 ``shower'' neutrino events above 60~TeV, 12 events are emitted with associated GRBs
by considering the Lorentz violation of neutrinos~\cite{Amelino-Camelia:2016ohi,Huang:2018}.
In this paper we make an estimate of the background of our previous analyzed
events and show that they are well stand out beyond the backgrounds. Further more,
we find that ``track" events can also be associated with GRBs
	under the same Lorentz violation features of neutrinos.
Our results of track events are consistent with the shower events with a same Lorentz violation scale $E_{\rm LV}$. These results can be considered as new supports to the obtained LV scale $E_{\rm LV}$ from
previous works~\cite{Amelino-Camelia:2016fuh,Amelino-Camelia:2016ohi,Huang:2018}. It is also indicated that the neutrino flux from GRBs associated by us is comparable with the prediction from GRB fireball models.	We therefore conclude that gamma ray bursts can serve as a significant source
for the ultra-high energy IceCube neutrino events
provided that there are Lorentz violation of neutrinos and anti-neutrinos.
Our work supports the Lorentz violation and $CPT$-violation of neutrinos as proposed in ref.~\cite{Huang:2018}, indicating new physics beyond relativity.

\section*{Declaration of Competing Interest}
The authors declare that they have no conflicts of interest in this
work.

\section*{Acknowledgements}

This work is supported by National Natural Science Foundation of China (Grant No.~12075003).

\end{document}